# SPECTRAL REFLECTANCE BASED HEART RATE MEASUREMENT FROM FACIAL VIDEO

*Arvind Subramaniam[1] and Rajitha.K[2]*
1. Department of Electrical and Electronics Engineering, BITS PILANI Hyderabad Campus, Telangana, India
2. Department of Civil Engineering, BITS PILANI Hyderabad Campus, Telangana, India

**ABSTRACT**

Remote detection of the cardiac pulse has a number of applications in sports and medicine, and can be used to determine an individual's physiological state. Previous approaches to estimate Heart Rate (HR) from video require the subject to remain stationary and employ background information to eliminate illumination interferences. The present research proposes a spectral reflectance-based novel illumination rectification method to eliminate illumination variations in the video. Our method does not rely on the background of the video and is robust to extreme motion interferences (head movements). Furthermore, in order to tackle extreme motion artifacts, the present framework introduces a novel feature point recovery system which recovers the feature tracking points lost during extreme head movements of the subject. Finally, the individual HR estimates from multiple feature points are combined to produce an average HR. We evaluate the efficacy of our framework on the MAHNOB-HCI dataset, a publicly available dataset employed by previous methods. Our HR measurement framework outperformed previous methods and had a root mean square error (RMSE) of 5.21%.

## 1. INTRODUCTION

Standard techniques to determine HR involve physical contact and are inconvenient for the user. Several papers have proposed heart rate measurement techniques which utilize facial video for the estimation of the cardiac pulse [1], [2]. Garbey et al. used the thermal signal emitted by facial blood vessels due to the influx of blood at every heartbeat [3]. Since the approach is extremely sensitive to thermal signals, an elevation in body temperature of the user will result in erroneous results. Moreover, it does not take illumination and motion artifacts into account.
In 2014, Li et al. proposed a method to tackle illumination changes using the background of the video [4]. Assuming equal spectral reflectance for the foreground and the background, they neutralized the variation in the foreground by subtracting the illumination variation of the background from the face of the user. Although the method had reasonably high accuracy, it is likely to fail in case the spectral reflectance of the background and skin are different. Huan et al. in 2017, deviated from the ICA-based approaches and employed Joint Blind source separation (JBSS) to measure HR [5]. Zhang et al. employed a six channel ICA algorithm to simultaneously detect HR and blink [6]. However, both these methods do not account for motion or illumination interferences. A spectral reflectance based approach was proposed by Lam et al. in 2015 [7], which used multiple HR estimates from local regions to estimate the final HR. However, there is a lot of computation involved due to which the overall speed is compromised. Moreover, HR is computed from arbitrarily chosen feature tracking points, which makes the estimate unreliable.

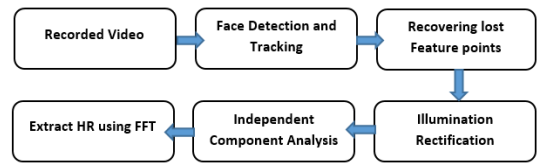

Figure 1: Stages of the proposed framework.

The proposed method avoids large computations by providing a simpler and more robust framework to measure HR from video.
We propose a novel HR measurement framework which consists of a feature point recovery system and a spectral reflectance-based illumination rectification approach to deal with motion and illumination interferences respectively. Our illumination rectification approach is independent of the background and produces excellent results even when the foreground and background have different illumination sources, unlike previous papers which have utilized the mean pixel intensity of the background to neutralize changes in illumination [4].
In Section 2, after performing face detection and tracking, we propose a feature point recovery system to deal with motion artifacts such as extreme head rotations. Section 3 introduces a novel approach to rectify changes in illumination. We primarily focus on green channel intensity in Section 3. The experimental results and comparison with previous approaches follow in Section 4. Details of the MAHNOB-HCI dataset are mentioned in Section 4.2. Finally, Section 5 draws conclusions.

## 2. RECOVERY OF FEATURE TRACKING POINTS

We employed the pose-free facial landmark fitting tracker for face detection and tracking [8]. This tracker has been employed by previous HR estimation methods and can simultaneously handle face detection, pose-free landmark localization and tracking over a large range of motions in real time [7].
Following face detection and tracking, a feature point recovery system has been devised to overcome extreme motion artifacts. In case of large head movements, it is possible for parts of the subject's face to get obscured from the camera. As a result, this may lead to the loss of a large number of feature points. In order to recover the lost feature points, we monitor the total number of features at any given instant and compare it with a threshold value. In case the number of tracking points falls below the threshold value, the landmark tracker is reapplied, exactly 36 frames (0.6s) after the frame where the failure had occurred. We have taken the threshold value to be 60% of the total number of feature points.

However, monitoring the total number of feature points is not a sufficient requirement. Even though reapplying the tracker would produce nearly the same number of feature points as before, it would not be possible to recover the feature points corresponding to the obscured portion of the face, since the new feature points would have different locations. For instance, in the central figure in Fig. 2, even though we have obtained the same number of feature points as before, the feature points corresponding to the obscured portion of the face are missing. To resolve this, we compute the root mean square error (RMSE) of the new feature point centroid ($\mu(t)$) relative to the old centroid ($\sigma(t)$) using

$$\epsilon(t) = \sqrt{\frac{\sum_{i=1}^{N}|\mu(t) - \sigma(t)|^2}{N}}, \quad (1)$$

where N denotes the number of features points at a given frame t and $\epsilon(t)$ denotes the RMSE. An important point to note is that the second step would take place only if the first criterion is satisfied. The landmark tracker is applied once every 36 frames (0.6 seconds) until $\epsilon(t)$ reaches a minima. Fig. 3 illustrates the choice of RMSE and the corresponding frame at which the feature points are recovered. Face detection and tracking are not re-implemented beyond this frame. Since the pose-free landmark tracker takes less than 0.20 seconds per image combined, our framework can tackle extreme head rotations without increasing the computational complexity or compromising the overall speed of the model. We have chosen a time interval of 0.6 seconds since the feature point recovery system takes a total of 0.35 seconds for face detection, tracking and computation of RMSE (maximum).

As shown in Fig. 2, after recovering lost feature points, it is possible to consistently track the subject's face and obtain the feature points despite considerable head movement. We have plotted the RMSE for a 90 degree rotation of the subject. As one may observe in Fig.4, the RMSE starts from zero since there is no deviation between the new and the old centroid initially. However, it gradually increases as the subject's head rotation increases. An interesting point to note is the absence of an RMSE point corresponding to frame 180 (after failure). This is because the number of feature tracking points are less than the threshold of 60%. Since the first condition has failed, the RMSE has not been calculated from frames 150 to 190 (when a portion of the face is completely obscured from the camera).

As shown in Fig. 3, the process of recovering feature tracking points is complete after the minima is achieved. Consequently, it terminates the process and moves on to the next step – Illumination Rectification.

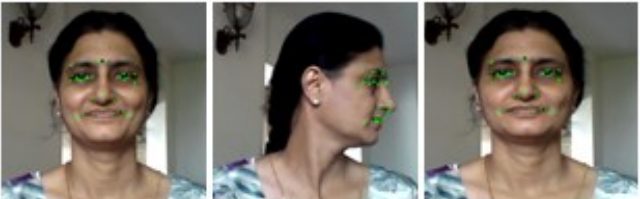

Figure 2: Result after recovery of feature tracking points. The features in the left part of the face are recovered, even after being completely obscured from the camera. From left to right, the frame numbers are 21, 170 and 309.

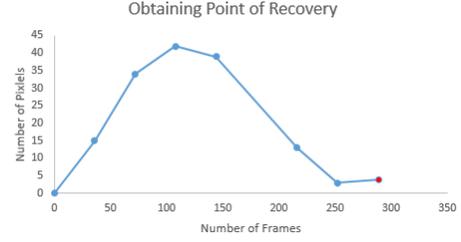

Figure 3: The RMSE represents the difference in the number of pixels between $\mu(t)$ and $\sigma(t)$. The recovery of feature points terminates after the minima is achieved – in this case, 288 frames (4.7 seconds) after tracking failure.

## 3. ILLUMINATION RECTIFICATION

The spectral reflectance of the skin is a result of two components: the influx of blood to the face at every heartbeat, and the pigmentation of the skin [9]. Hence, we have modeled the intensity of a facial pixel as:

$$I(t) = I_b(\rho, t) + I_p(\rho, t) \quad (2)$$

where $I_b$ and $I_p$ are the pixel intensities due to blood flow and the pigmentation (due to melanin, carotenes, etc.) of the subject. The intensity can be represented as:

$$I(t) = \int R(\rho, t) P(\rho) C_i(\rho) d\rho \quad (3)$$

where $P(\rho)$ is the normalized spectral power distribution of the illuminant, $C_i(\rho)$ is the camera spectral sensitivity for channel $i$ and $R(\rho, t)$ is the spectral reflectance of the point over wavelengths $\rho$ [10]. However, as a result of illumination variations, the intensities in eqn. 2 have a component corresponding to illumination interferences as well. Hence, combining eqns. 2 and 3, the total intensity obtained from the face of the subject is:

$$I(t) = I_{b(impure)}(\rho, t) + I_{p(impure)}(\rho, t) \quad (4)$$

where $I_{b(impure)}(\rho, t) = \int R_b(\rho, t) P(\rho) C_i(\rho) d\rho$ is the intensity due to blood flow (with illumination variations), and $I_{p(impure)}(\rho, t) = \int R_p(\rho, t) P(\rho) C_i(\rho) d\rho$ represents the intensity corresponding to pigmentation (chiefly due to melanin and carotenes) respectively. An important point to note is that these intensities consist of illumination interferences as well. After obtaining the two intensities using independent component analysis (ICA), we utilize the illumination interferences in $I_{p(impure)}(\rho, t)$ to remove the illumination interferences in $I_{b(impure)}(\rho, t)$.

Since the radiance and the corresponding illumination variations are consistent within the subject's face, the spectral distributions corresponding to the illumination variations of skin pigmentation would be equal to that of blood flow. Our illumination rectification approach is similar to Li et al. [4]. However, we do not take the background into account and instead, employ spectral reflectance to get the intensities due to blood flow and pigmentation. The intensity due to pigmentation acts as the background, using which we are able to cancel the illumination interferences in $I_{b(impure)}(\rho, t)$. Similar to Li2014, we model illumination interferences as a linear function of $I_{p(impure)}(\rho, t)$.

$$I_{IV}(\rho, t) \approx K I_{p(impure)}(\rho, t) \quad (5)$$

where $I_{IV}(\rho, t)$ represents intensity solely corresponding to illumination interferences. Hence, $I_{b(impure)}(\rho, t)$ can be modeled as:

$$I_{b(impure)}(\rho, t) = I_{b(pure)}(\rho, t) + I_{IV}(\rho, t) \quad (6)$$

From eqns. 5 and 6, we have:

$$I_{b(pure)}(\rho, t) = I_{b(pure)}(\rho, t) - K I_{p(impure)}(\rho, t) \quad (7)$$

Rather that solely looking at eqn.7, we have focused on eq.5 and tried to minimize the error $I_{IV}(\rho, t) - K I_{p(impure)}(\rho, t)$. Once the optimal K is estimated, the reasonable approximation for $I_{IV}(\rho, t)$ can be computed.

We have utilized the Recursive Least Squares (RLS) adaptive filter to calculate the ideal value of K that minimizes the error. The RLS adaptive filter is an algorithm that recursively computes the filter coefficients that minimize a linear cost function related to the input signal [11].

Let $K(t)$ be the estimated filter weight for each point time point $t$. After initializing the weights, the RLS filter updates the filter weights as

$$K(t+1) = K(t) + C^{-1}(t) I_{b(pure)}(\rho, t) I_{p(impure)}(\rho, t) \quad (8)$$

Here, $C(t)$ is the autocorrelation matrix given by

$$C(t) = \sum_{i=0}^{t} I_{b(pure)}(\rho, t) I_{p(impure)}^{T}(\rho, t) \alpha^{t-1}, \quad (9)$$

where $I_{p(impure)}^{T}(t)$ is the transpose of $I_{pimpure}(t)$ and $\alpha$ is a positive constant smaller than 1. The RLS filter will continue to run its iterations until K(t) converges to a suitable value that minimizes the error/deviation $I_{IV}(\rho, t) - K I_{p(impure)}(\rho, t)$. We have applied the Local Region Based Active Contour (LRBAC) method to segment the background region of each frame [12]. Since LRBAC is a region-based approach, it is insensitive to image noise, as opposed to edge-based methods such as the Distance Regularized Level Set Evolution (DRLSE) method used by Li et al. Fig. 4 shows the green channel skin pigmentation intensity before and after illumination rectification. As one may observe, the peaks representing illumination interferences have been removed without distorting intensity values corresponding to normal illumination. We have also compared the illumination rectified green channel blood flow intensity with that of the ground truth signal in Fig. 5. As it can be seen, the illumination rectified intensity we obtain is almost identical to the ground truth signal.

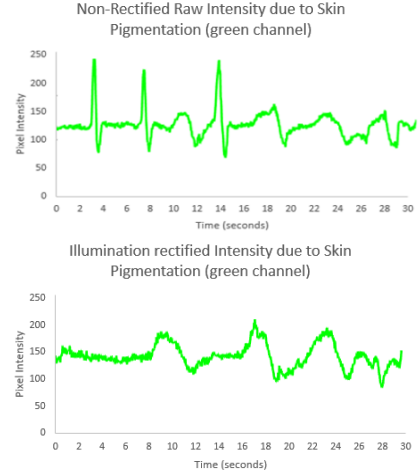

Figure 4: The skin pigmentation intensity corresponding to the spectral reflectance of the skin (green channel). The illumination interferences (peaks at 3s, 8s and 14s) have been removed post rectification (bottom).

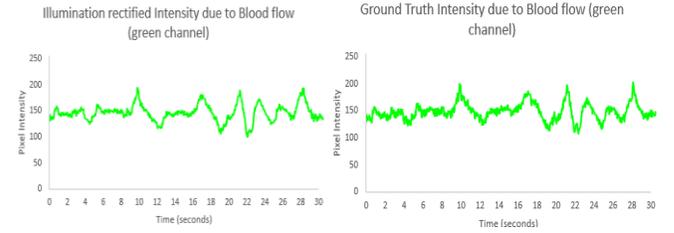

Figure 5: The illumination rectified signal due to spectral reflectance of blood flow (green channel) and the ground truth green pixel intensity (left). Ground truth is obtained from the recording of the same scene in conditions without illumination variations (right).

## 4. RESULTS

### 4.1. HR Extraction

Following illumination rectification, we normalize each channel (RGB) and apply ICA to extract plethysmographic (PPG) signals. After performing FFT on each PPG signal, we found that the PPG signal corresponding to the green colour channel had the most prominent frequency among the three colour channel, as shown in Fig. 6. Finally, temporal filters are applied to band limit the range of frequencies corresponding to the HR of an average individual. As opposed to computing HR from the mean facial pixel intensity, illumination rectified pixel intensities of selective feature points are obtained. This enables a more accurate and reliable HR estimate since individual feature points can be chosen based on their location, unlike a random selection of feature points as done by Lam et al. Since the HR of a person ranges from 40-240 bpm, we set the range of frequency as [0.6, 4] Hz. In this instance, the HR was found to be 1.21 Hz, which is equivalent to 72.6 bpm. An important to note is that the HR is obtained from feature points located in the central portion of the face (nose). This is to ensure that our HR estimate is free from interferences such as facial hair, spectacles, sunglasses, etc. For instance, the PPG signal obtained from the eye or cheek of the subject may be distorted due to the afore-mentioned interferences. After choosing feature points in the central portion of the subject's face, multiple HR estimates from each feature point have been combined to produce a single average HR. Fig. 6 directly shows the final average HR obtained. The process of combining individual HR estimates makes the system tolerant to noise and ensures that the framework works well even if the illumination interferences within the subject's face are different. In other words, our framework does not assume uniform blood flow and illumination across a subject's face. This adds a greater degree of robustness to our method.

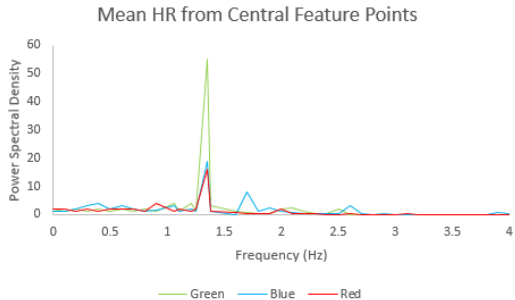

Figure 6: Mean HR obtained from feature points in the central portion of the subject's face.

### 4.2. Performance on MAHNOB-HCI dataset

The MAHNOB-HCI database is a public database comprising 527 colour videos of 27 subjects (12 males and 15 females). The videos have a resolution of 780 x 580 pixels recorded at 61 FPS. In order to be consistent with previous methods, we chose videos from the emotion elicitation portion of the database. Since the videos were of varying lengths, we used frame 306 to frame 2135 (30 seconds) of each video to measure the average heart rate.

In Table 1, [i], [ii], [iii] and [iv] refer to face detection and tracking, Feature point recovery, Illumination rectification and Temporal filtering respectively.

| Framework | RMSE (%) | MAE | % Absolute error < 5 bpm | r |
|---|---|---|---|---|
| Huan2017 | 14.5 | 8.9 | 59.7 | 0.63* |
| Zhang2017 | 15.7 | 8.7 | 63.2 | 0.65* |
| Li2014 | 15 | 7.8 | 68.1 | 0.728* |
| Lam2015 | 8.9 | 4.7 | 75.1 | 0.85* |
| Ours ([i]+[iv]) | 28.2 | 16.5 | 42.1 | 0.31 |
| Ours([i]+[ii]+[iv]) | 19.2 | 11.3 | 56.3 | 0.59* |
| Ours([i]+[iii]+[iv]) | 16.8 | 10.8 | 61.2 | 0.58* |
| **Ours (All Steps) ([i]+[ii]+[iii]+[iv])** | **5.21** | **3.5** | **86.4** | **0.91*** |

Table.1. Performance and comparison of each framework on the MAHNOB-HCI dataset. * indicates that the result is statistically significant at p = 0.01.

We consider the RMSE, Mean Absolute Error (MAE), correlation r and the percentage of Absolute Error less than 5bpm. As it can be seen from Table.1, our method (after including all steps) outperforms all the other methods, including Zhang2017 and Huan2017, since these models do not account for illumination interferences. Zhang 2017 performed slightly better than Huan2017, as a result of using a six channel ICA, which automatically eliminated minor motion and illumination artifacts through one of the six channels (other than the two channels corresponding to BVP and blink rate). While Lam2015 had the best results among previous methods, their framework chose arbitrary feature points to calculate HR, which may compromise the accuracy in case the chosen feature points are unsuitable and can provide a distorted blood flow intensity.

In addition to evaluating the performance of previously proposed methods, we have also evaluated the contribution of each component of our framework in Table 1. We obtained an increase in correlation r from 0.31 to 0.91 upon inclusion of (ii) and (iii) in our framework. Our method outperforms Li2014 since the latter assumes equal spectral reflectance for the foreground and the background which may not always be the case. Further, we also compare the HR error of our framework with Li2014 and Lam2015. As shown in Fig.7, we estimate 421 cases (86.4%) with errors of less than 5 bpm, while the numbers for Li2014 and Lam2015 are 332 (68.1%) and 366 (75.1%) respectively.

An important point to note is that not all videos could be used since the ground truth data was unavailable for certain videos. Hence, we were ultimately able to use 487 videos. As a result, our implementation of Li2014 has a different result than the original paper, since it was tested on 527 videos. However, our results concur with Lam2015, since the latter was also implemented on 487 videos in the MAHNOB-HCI dataset.

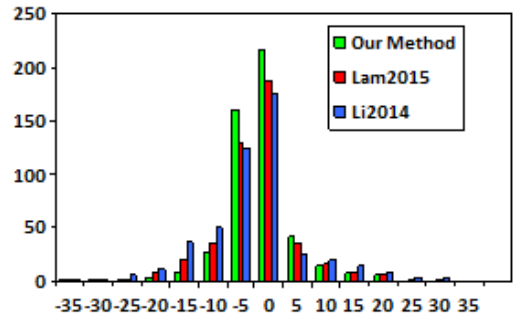

Figure 7: Comparing HR error distributions of our method with Li2014 and Lam2015

### 5. CONCLUSION

In this paper, we have proposed a novel feature point recovery system to tackle motion artifacts and an illumination rectification method to address the problem of illumination variations. An important aspect of our framework is that our approach does not depend on the background to rectify illumination artifacts. As a result, we obtain a higher accuracy on the MAHNOB-HCI dataset as compared to previously proposed methods. Our approach ensures accurate HR measurement even if the background and foreground are illuminated by different light sources. Next, instead of relying on a single HR estimate, we utilized independent HR estimates from feature points in the central portion of the subject's face – which is relatively free from interferences such as facial hair, spectacles, etc. – to recover a single average HR estimate. This enables our framework to have a greater degree of robustness to illumination interferences.

A direction for future work would be to focus on the application of accurate video-based HR measurement in medical diagnosis such as the detection of potentially serious breathing disorders like sleep apnea.